\documentstyle[prl,twocolumn,aps,english]{revtex}

\def\beq{\begin{equation}}
\def\eeq{\end{equation}}

\begin{document}

\title{The speed of light need not be constant}

\author{Asher Peres}
\address{Department of Physics, Technion---Israel Institute of
Technology, 32000 Haifa, Israel}

\maketitle
\begin{abstract}
Recent observations of the fine structure of spectral lines in the early
universe have been interpreted as a variation of the fine structure
constant. From the assumed validity of Maxwell equations in general
relativity and well known experimental facts, it is proved that $e$
and $\hbar$ are absolute constants. On the other hand, the speed of
light need not be constant.

\end{abstract}

\bigskip

There have recently been indications that the fine structure of
spectral lines in the early universe differs from its usual properties
\cite{webb}.  This has been interpreted as a slow change in the fine
structure constant over cosmological times, and there is a lively
controversy \cite{duff} whether this is due to variations of $e$,
$\hbar$, or $c$. In this note it is proved, from the validity of
Maxwell equations in general relativity and well known experimental
facts, that $e$ and $\hbar$ are absolute constants. On the other hand,
the speed of light need not be constant.

It is often stated that the speed of light is exactly 299\,792\,458
m/sec \cite{codata}, owing to a 1983 decision of Comit\'e International
des Poids et Mesures on how to define the unit of length ``meter''
\cite{giacomo}. However, this administrative decision raises serious
issues \cite{petley,nature}. Such a way of defining length implies
that there must be {\it synchronized\/} clocks at the extremities
of the measured object in order to determine the corresponding time
interval. Synchronization is best performed by light signals \cite{1905}
so that we effectively have a circular definition. Einstein himself
wrote, after the first equation of his historic paper, ``we assume
that this definition of synchronization is free from contradictions,
and possible for any number of points.'' Contradictions naturally appear
in non-inertial coordinate systems. For example, if we are serious in
measuring distance by means of the elapsed time for light signals,
then the circumference of the Earth at the equator is 123.9m longer
if measured eastward than if measured westward \cite{nature}.

Genuine gravitional fields cause even more difficulties. Again quoting
Einstein \cite{1912}: ``the constancy of the velocity of light can be
maintained only insofar as one restricts oneself to spatio-temporal
regions of constant gravitational potential.'' Serious practical
difficulties indeed have to be overcome for setting the global positioning
system (GPS) \cite{gps}. In summary, the legal value of $c$ is not in
general the actual speed of electromagnetic signals. The only meaning of
this exact, invariant, official speed of light is to serve for writing
Lorentz transformations, if there is any need of them.

Let us briefly examine some alternatives. Can $\hbar$ change?  Planck's
constant is more than just a conversion factor between Joules and Hertz.
Quantum systems are not localized, they are pervasive. In particular,
entangled systems may be spread over arbitrary distances.  A value
of $\hbar$ varying in spacetime would necessitate a complete revision
of quantum theory.  Can the electric charge vary? It is a historical
accident that Coulomb's law of force between macroscopic charges
was discovered before it was known that the electric charges of
all particles are integral multiples of $e$ (or $e/3$ if we include
quarks). This indicates that we should define $e=1$ as the unit of
charge (this is a {\it natural\/} unit, not a conversion factor). A
more formal proof of constancy will be given below.

We must therefore have a closer look at $c$. It cannot be a mundane
conversion factor. Time is not a fourth dimension of space. Relativistic
transformations never change the nature of timelike and spacelike
intervals; important information is lost if we ignore the difference.
Likewise, in the gravity field of the Earth, the vertical and horizontal
directions are not equivalent. Airlines measure them in feet and miles,
respectively. The use of different length units may be advantageous in
dimensional analysis: it is easily seen that a trajectory with initial
velocities $(v_x,v_h)$ must be a parabola $h=av_h(x/v_x)+bg(x/v_x)^2$.
(A complete dynamical calculation gives $a=1$ and $b=-1/2$.)

In general relativity, all four coordinates may have different
dimensions and the constant $c$ does not appear at all in the
fundamental equations (it may appear in particular solutions, once
sources that are not generally covariant have been specified with
arbitrary units). In the early universe, where background radiation
cannot be ignored, Lorentz invariance does not hold. There is a
preferred frame. It is then plausible that, in such an environment,
the ``vacuum'' behaves as a dielectric medium where the speed of light
is different from its ideal value \cite{ll}.

In this paper, I shall not use the full machinery of general relativity,
but only the invariance of the fundamental equations under arbitrary
nonlinear coordinate transformations. Einstein's gravitational equations
are not used, Newton's constant $G$ does not appear, so that Planck
units, whatever they mean, are not involved. The only assumption is the
validity of Maxwell's equations

\beq F_{\mu\nu,\rho}+F_{\nu\rho,\mu}+F_{\rho\mu,\nu}=0, \label{F}\eeq
and

\beq {\cal F}^{\mu\nu}{}_{,\nu}={\cal J}^\mu, \label{J} \eeq
where the covariant antisymmetric tensor $F_{\mu\nu}$ corresponds to the
field components usually called {\bf E} and {\bf B}, and the
contravariant antisymmetric tensor density ${\cal F}^{\mu\nu}$
corresponds to {\bf D} and {\bf H}. Commas denote partial derivatives
(not covariant derivatives) and ${\cal J}^\mu$ is a vector density.
The existence of a metric is not needed for the above form of Maxwell's
equations. Lorentz invariance is irrelevant to them. The metric appears
only in the relation between $F$ and $\cal F$ which is, in vacuo,

\beq {\cal F}^{\mu\nu}=
  \sqrt{-g}\,g^{\mu\rho}\,g^{\nu\sigma}\,F_{\rho\sigma}.\label{G} \eeq

This form of Maxwell equations was first written by Einstein
\cite{1916}. Later, they were again derived by Weyl \cite{weyl} and
Brillouin \cite{brillouin} who systematically developed the notion of
densities and capacities. It has even been proposed to consider $F$ and
$\cal F$ as the fundamental fields, and the metric, defined by
Eq.~(\ref{G}), as a derived quantity \cite{aop,obukhov,hehl}.
 
An apparent difficulty, which will actually turn into a powerful tool,
is the fact that spacetime coordinates may have different dimensions
(time, length, angles, and so on). Then likewise the components of $F$
have different dimensions, and also those of $\cal F$. The notion of
{\it absolute dimension\/} of a tensor was introduced by Schouten
\cite{schouten} and Post \cite{post}. As a simple example, consider the
electric charge

\beq Q=\int{\cal J}^\mu\,dS_\mu, \eeq
where $dS_\mu$ is an element of a three-dimensional hypersurface, such
as $dxdydz$ or $drd\theta d\phi$. Therefore $dS_\mu$ transforms as a
covariant vector capacity, so that $Q$ is invariant under arbitrary
coordinate dimensions, and we say that the absolute dimension of
$\cal J$ is [Q]. It is also the absolute dimension of $\cal F$,
owing to Eq.~(\ref{J}). It follows that the electronic charge $e$
is invariant, as we already found by a more intuitive reasoning.

As we have seen, dimensional analysis is an important tool for studying
nature and deriving useful results. Dimensional analysis also applies
to absolute dimensions: they must be the same on both sides of
any equation.  Consider now Eq.~(\ref{F}). It states that there are
no magnetic charges and guarantees the local existence of a covariant
vector field $A_\mu$ such that

\beq F_{\mu\nu}=A_{\mu,\nu}-A_{\nu,\mu}. \eeq
For any given closed path, the value of the magnetic flux,

\beq \Phi=\oint A_\mu\,dx^\mu, \eeq
is a scalar (i.e., it is invariant under any coordinate
transformations). It follows that the absolute dimension of $F$ is that
of magnetic flux, namely [action/charge]. Now, it is experimentally
known that in some superconductors \cite{degennes}, flux appears
in integral multiples of $h/2e$. This is the natural unit of flux,
just as $e$ is the natural unit of charge. It follows that $\hbar$
is an absolute constant, in agreement with the intuitive argument
presented above.

There is no similar argument for the constancy of $c$ (or for mass
ratios of elementary particles, and other important ``constants''). It
is plausible that in the early universe, when matter and radiation
were much more concentrated than today, atomic energy levels were
different from those of truly isolated atoms, in a way analogous to
the chemical shift of nuclear energy levels in the M\"ossbauer effect
\cite{mossbauer}.  However this phenomenon is not yet understood, and
more work, theoretical and experimental, is needed.  We may then hope
that more precise observations on the fine structure of spectral lines
will give information on the properties of the cosmic environment in
early universe.

\bigskip I am grateful to Friedrich Hehl and Rafael Porto for many
helpful comments on my preceding communication \cite{qph}. This work
was supported by the Gerard Swope Fund and the Fund for Promotion
of Research.

\end{document}